\documentstyle[sprocl,epsfig]{article}

\def\nostrocostrutto#1\over#2{\mathrel{\mathop{\kern 0pt \rlap 
  {\raise.2ex\hbox{$#1$}}}
  \lower.9ex\hbox{\kern-.190em $#2$}}}
\def\gsim{\nostrocostrutto > \over \sim}   

\newcommand{\be}{\begin{equation}}
\newcommand{\ee}{\end{equation}}
\newcommand{\ba}{\begin{eqnarray}}
\newcommand{\ea}{\end{eqnarray}}

\catcode`@=11
\newcount\@tempcntc
\def\@citex[#1]#2{\if@filesw\immediate\write\@auxout{\string\citation{#2}}\fi
  \@tempcnta\z@\@tempcntb\m@ne\def\@citea{}\@cite{\@for\@citeb:=#2\do
    {\@ifundefined
       {b@\@citeb}{\@citeo\@tempcntb\m@ne\@citea\def\@citea{,}{\bf ?}\@warning
       {Citation `\@citeb' on page \thepage \space undefined}}%
    {\setbox\z@\hbox{\global\@tempcntc0\csname b@\@citeb\endcsname\relax}%
     \ifnum\@tempcntc=\z@ \@citeo\@tempcntb\m@ne
       \@citea\def\@citea{,}\hbox{\csname b@\@citeb\endcsname}%
     \else
      \advance\@tempcntb\@ne
      \ifnum\@tempcntb=\@tempcntc
      \else\advance\@tempcntb\m@ne\@citeo
      \@tempcnta\@tempcntc\@tempcntb\@tempcntc\fi\fi}}\@citeo}{#1}}
\def\@citeo{\ifnum\@tempcnta>\@tempcntb\else\@citea\def\@citea{,}%
  \ifnum\@tempcnta=\@tempcntb\the\@tempcnta\else
   {\advance\@tempcnta\@ne\ifnum\@tempcnta=\@tempcntb \else \def\@citea{--}\fi
    \advance\@tempcnta\m@ne\the\@tempcnta\@citea\the\@tempcntb}\fi\fi}
\catcode`@=12

\begin{document}

\setcounter{page}{0}
\thispagestyle{empty}

\noindent
 \rightline{MPI-PhT/97-40}
 \rightline{MEIJI/THEP/7/97}
 \rightline{May 30, 1997}
\vspace{0.5cm}
\begin{center}
{\Large \bf Rapidity Gaps in $e^+e^-$ Annihilation \\ 
and Parton-Hadron Duality}
\end{center}
\vspace{1.0cm} 
\begin{center}
WOLFGANG OCHS~\footnote{E-mail: wwo@mppmu.mpg.de}  
\end{center}

\begin{center}
\mbox{ }\\
 {\it Max-Planck-Institut f\"ur Physik \\
(Werner-Heisenberg-Institut) \\
F\"ohringer Ring 6, D-80805 Munich, Germany} 
\end{center}

\begin{center}
\mbox{ }\\

and

\mbox{ }\\
TOKUZO SHIMADA~\footnote{E-mail: tshimada@gravity.mind.meiji.ac.jp}
\end{center}

\begin{center}
\mbox{ }\\
{\it Meiji University, Department of Physics\\
Higashi Mita 1-1, Tama\\
Kawasaki, Kanagava 214, Japan} 
\end{center}

\vspace{0.5cm}

\begin{abstract}
We calculate the probability for rapidity gaps in the parton cascade for
different approximations within the perturbative QCD and compare 
the results with recent measurements. The aim is to find out whether 
the  dual connection between the
parton and hadron final states -- observed so far in various inclusive
measurements -- holds as well for the extreme kinematic
configurations with colour sources separated by  large rapidity gaps.
A description of the data is possible indeed choosing  the 
parameters of the cascade in the range 
suggested by recent analyses of the energy spectra
(the $k_\perp$ cutoff $Q_0~\gsim$ QCD 
scale $\Lambda\sim 250$ MeV).
\end{abstract}

\vfill 
\noindent 
to appear in the Proc. of the 33rd Eloisatron Workshop 
``Universality features in multihadron production and the leading effect'', 
(Erice, Italy, 19-25 October 1996) 

\newpage 

\title{RAPIDITY GAPS IN $e^+e^-$ ANNIHILATION \\
AND PARTON-HADRON-DUALITY}

\author{WOLFGANG OCHS}

\address{Max-Planck-Institut f\"ur Physik (Werner-Heisenberg-Institut) \\
F\"ohringer Ring 6, D-80805 Munich, Germany\\ 
E-mail: wwo@mppmu.mpg.de}

\author{TOKUZO SHIMADA}

\address{Meiji University, Department of Physics\\
Higashi Mita 1-1, Tama\\
Kawasaki, Kanagava 214, Japan\\
E-mail: tshimada@gravity.mind.meiji.ac.jp}

\maketitle
\abstracts{
We calculate the probability for rapidity gaps in the parton cascade for
different approximations within the perturbative QCD and compare
the results with recent measurements. The aim is to find out whether
the  dual connection between the
parton and hadron final states -- observed so far in various inclusive
measurements -- holds as well for the extreme kinematic
configurations with colour sources separated by  large rapidity gaps.
A description of the data is possible indeed choosing  the
parameters of the cascade in the range
suggested by recent analyses of the energy spectra
(the $k_\perp$ cutoff $Q_0~\gsim$ QCD
scale $\Lambda\sim 250$ MeV).}

\section{Introduction}

Recently the studies of $p \bar p$ collisions with high $p_T$ jets at
the Tevatron 
\cite{teva1,teva2} and of dijet events in $ep$ collsions at HERA
\cite{hera1,hera2}
showed that a considerable fraction of these hard scattering events
(of order 1-10\%) contain large gaps in the rapidity distribution of
particles. Such phenomena had been expected from the exchange of
colour neutral objects \cite{bj}.

In $e^+e^-$ annihilation the primary parton process is the production of 
$q \bar q$ pairs followed by gluon bremsstrahlung and quark pair production.
In this case the suppression of central gluon radiation would
produce a rapidity gap for the partons but the recoiling $q \bar q$ like 
systems separated by the gap may become the source of nonperturbative
hadron production to neutralize the color field; then 
no gap would remain between the observed hadrons.

One way to generate large rapidity gaps between the final hadrons
in $e^+e^-$ annihilation is through the production of 
colour singlet clusters separated by large rapidity gaps. Such
configurations can be obtained perturbatively in hard processes
leading to at least four partons in the final state,
like $e^+ e^- \to q\bar q q\bar q$ or $q\bar q gg$
\cite{bbh,er}.
In this case the hadrons are produced within each colour neutral 
$q\bar q$ or $gg$ cluster and the
large gap persists.
As these processes involve a highly virtual intermediate quark or gluon 
the rate for these large rapidity gap events is 
rather small though. One finds that the rate keeps decreasing 
with increasing gap size, contrary to the case of $p \bar p$ or $ep$ collision.

Such an unlimited decrease of the gap probability has been seen
already some time ago \cite{hrs}.
The recent analysis at the $Z^0$ resonance by the SLD Collaboration
\cite{sld} has  shown the fall of the gap probability over
five orders of magnitude (see also the first results from
ALEPH \cite{aleph}).
At the same time the observed absolute size of the gap probability
exceeds the expectations from the above calculations
 \cite{bbh,er} by about two orders
of magnitude. This observation excludes this type of hard colour 
neutralization 
to be the dominant mechanism for the production of large rapidity gaps.

In this paper we investigate how the fraction of rapidity gap events
$f(\Delta y )$ for the parton cascade (without hadronization) compares with the
experimental results on hadrons. At first sight one might
expect that the decrease of the gap fraction for hadrons and therefore 
in the experimental data is steeper
than the one of the parton cascade because of the additional
suppression from removing the hadrons emitted into the gap
during the hadronization phase. We will see however that this effect
strongly depends on the scale $Q_0$ for which the parton cascade is
terminated.

There are many phenomena in multiparticle production which are in
favour of a very simple hadronization picture, called local parton hadron
duality (LPHD) \cite{lphd}, which compares the observed sufficiently
inclusive particle densities and correlations directly with the
calculations at the parton level, assuming in effect the production
of a hadron close to the parton in phase space (for reviews, see
\cite{dkmt,ko}).
In this approach the parton cascade is evolved down to the low scales
$Q_0 \gsim \Lambda$ of a few hundred  MeV; $Q_0$ denotes here the cutoff in
the transverse momentum of the emitted partons
and $\Lambda$ the QCD scale for the running coupling.

This calculation of the gap rate is similar in spirit to the
``clan'' model analysis within a simplified parton shower interpretation:%
\cite{clan} the primary independent emission of hadronic objects
leads to a Poisson distribution with the gap rate related to the
probability for no emission by an exponential.

In the following we calculate the gap probability analytically in
different approximation schemes of
perturbative QCD and compare to the data for the previously determined values
of $Q_0$ and $\Lambda$. This will teach us whether the parton hadron
duality works  in the more extreme, quasi exclusive configurations
with large rapidity gaps as well; in this case a confirmation of the
duality  approach 
is even less expected than in case of inclusive observables.

\section {Rapidity gaps in the parton cascade}

A gap in the rapidity distribution 
occurs if 
the initial $q \bar q$ pair does not radiate into the respective rapidity
interval.  
The probability for no radiation into a certain angular interval
is given in field theory by the
exponential Sudakov form-factor \cite{suda}, originally derived in QED.
A problem very similar to the rapidity gaps considered here 
 occurs in the calculation of multijet rates
in $e^+e^-$ annihilation applying the Durham/$k_\perp$
algorithm:\cite{cdotw,do}  
the 2-jet rate is again given by the Sudakov
form factor  which represents the probability for 
 no-parton-production above a certain
$k_\perp$ cutoff (resolution parameter).

In our application we consider the rapidity gap having no gluons inside the
angular or rapidity interval
above the transverse momentum cutoff $Q_0$ which corresponds to a hadronic
scale of a few hundred MeV as outlined in the introduction. 

Let us consider specifically the angular interval between $\Theta_1$ and $
\Theta_2$ ($\Theta_1 < \Theta_2$) where $\Theta$ is 
measured with respect to the quark direction. The rapidity is then obtained
from $y = -\ln {\rm tg} \frac{\Theta}{2}$. 
Let us further denote the probability for emission of a gluon
at an angle $\Theta^\prime$ with the energy $\omega^\prime$
off a parent parton $A$ (either a gluon(g) or a quark(q))
as                        
$ \wp_A(\omega^\prime,\Theta^\prime)=
dn_A/d\omega^\prime d\Theta^\prime$.

Then, the Sudakov form factor for the 
angular ordered cascade is given by
\begin{eqnarray}          
\Delta_A(P, \Theta,Q_0) &=& \exp (-w_A(P,\Theta,Q_0)) \label{Delta} \\
w_A(P,\Theta,Q_0) &=& \int d\omega^\prime \int_{k_\perp>Q_0} d\Theta^\prime
\wp_A(\omega^\prime,\Theta^\prime)
\label{exactdoubleintegration},
\end{eqnarray}            
and this represents the probability that no gluon is emitted
within the cone of half angle $\Theta$ from the parent parton
with the energy $P=Q/2$ with transverse momentum above $Q_0$.      
In particular,
\begin{eqnarray}
\Delta_A(\Theta_2)/\Delta_A(\Theta_1)
= \exp \left( -w_A(\Theta_2)+w_A(\Theta_1) \right)
\end{eqnarray}
represents the probabilty that there is no emission of a gluon
with emission angle between $\Theta_1$ and $\Theta_2$.

These rates have been calculated in
different approximations which we discuss in the following. 
\\

\noindent
{\it The double logarithmic approximation (DLA)}\\
The simplest approximation takes into account only the leading
contributions from the angle and energy singularities of the gluon emission. 
In this case the total number of gluons radiated from a  primary
parton into a forward cone of half angle
$\Theta $ as in (\ref{exactdoubleintegration}) is given 
in a small angle approximation by
\begin{equation}
w_A(P,\Theta,Q_0) = \frac{C_A}{N_C}
 \int_{\frac{Q_0}{\Theta}}^P 
                \frac{d\omega^\prime}{\omega^\prime} 
                \int_{\frac{Q_0}{\omega^\prime}}^\Theta 
                \frac{d\Theta^\prime}{\Theta^\prime}
\gamma_0^2(\omega^\prime\Theta^\prime)
       \label{wap}
\end{equation}
\noindent
or, using logarithmic variables 
\begin{equation}
w_A(Y,\lambda) = \frac{C_A}{N_C} \int_0^Y d\eta
     \int_0^\eta d\eta^\prime \gamma_0^2(\eta^\prime)
  \label{way}
\end{equation}
\noindent
with
\begin{equation}
Y=\ln\frac{P\Theta}{Q_0}, \qquad \lambda=\ln\frac{Q_0}{\Lambda}
                                  \label{logvar}
\end{equation}
\noindent
and the DLA anomalous dimension
$\gamma_0^2$ = $2N_C \alpha_s/\pi$
for running coupling $\alpha_s$
\begin{equation}
\gamma_0^2(\eta)=\frac{\beta^2}{\eta+\lambda},\quad 
    \beta^2=\frac{4N_C}{b}, \quad
    b=\frac{11}{3}N_C-\frac{2}{3}n_f
                                  \label{gamma}
\end{equation}
\noindent
where $N_C$ and $n_f$ denote the numbers of colours and flavours
respectively. Furthermore $C_A = \frac{4}{3}$ for quarks and 
 $C_A = N_C=3$ for gluons. One finds
\begin{equation}
w_A(Y,\lambda)=\frac{C_A}{N_C}\beta^2 
\{
(Y+\lambda)\ln \frac{Y+\lambda}{\lambda}-Y
\}
  \label{wdla}
\end{equation}
The probability for a gap without primary gluons 
is then obtained as
\begin{equation}
f_A(\Theta_1,\Theta_2)=e^{-(w_A(\Theta_2)-w_A(\Theta_1))}      \label{fth}
\end{equation}
\noindent
for two arbitrary angles $\Theta_1 < \Theta_2 < \frac{\pi}{2}$ measured 
with respect to the quark direction. In this approximation the jets evolve
independently in both hemispheres and the respective probabilities
factorize. In particular, in 
the case of a symmetric angular interval $(\Theta_G, \pi-\Theta_G$)
the gap probability is given by
\begin{equation}
f_A(\Theta_G)=e^{-2(w_A(\frac{\pi}{2})-w_A(\Theta_G))}   \label{fsym}
\end{equation}

In the present small angle approximation 
the rapidity is related to the angle by 
 $y \approx -\ln \frac{\Theta}{2}$; 
in case of the symmetric rapidity interval of full width
$\Delta y$, the relevant gap angle is $\Theta_G = 2 \exp (- \Delta y/2)$
which yields $\Theta_G = 2$ for the limiting case $\Delta y = 0$.
The use of the small angle approximation for such large angles is formally
not allowed, but we will see below that the effect is numerically small
(radiation occurs dominantly at small angles).
The gap probability $f_A(\Delta y)$ following from (\ref{fsym})
corresponds to an almost exponential decrease for the relevant parameter
range.

It is interesting to note that for fixed coupling there would be a
flattening of the gap fraction $f_A(\Theta_G)$ 
for large gaps $\Delta y$ as can be derived from
\begin{equation}
w_A(Y)=\frac{1}{2}\gamma_0^2Y^2\qquad ({\rm fixed }\alpha_s)
                                  \label{wfix}
\end{equation}
This difference comes from the strong radiation near the kinematical
limit $k_\perp \sim Q_0 \gsim \Lambda$ 
in case of running $\alpha_s$ which becomes
important for small limiting angles $\Theta $. In other words,
for large gaps the strong radiation near the kinematic limit 
is kinematically allowed and has
to be suppressed which makes the gap probability much smaller than
in the case of fixed $\alpha_s $ where this radiation is not
equally strong.\\

\noindent
{\it The modified leading logarithmic approximation (MLLA) }\\
Besides the double logarithmic terms one also 
takes into account in this approximation 
the leading single logarithmic corrections.
In this subsection we first derive the full O($\alpha_s$) emission probability
for $e^+e^-$ annihilation from which the various approximations beyond DLA
can be derived. 

We denote  the energy fractions
of $q,\bar q$ and $g$ as $x_+,x_-$ and $x_g$ and take the 
jet (thrust) axis along the (say, left-moving) $\bar q$. The energy
distribution is then given by 
\begin{equation}
 dn =  C_F \frac{ \alpha_s}{2 \pi}   
 \frac{x_{+}^2+x_{-}^2}{(1-x_{+})(1-x_{-})} dx_{+}dx_{-}.
 \label{exactwpinx+x-}
\end{equation}
(see, for example, Ref. \cite{hm} where also useful kinematic relations are
given).

In order to evaluate the gluon emission probablity from a parent
  quark
\begin{eqnarray}
\wp_q(\omega^\prime,\Theta^\prime) \equiv
\frac{ \partial(x_{+},x_{-}) } 
{ \partial(\omega^\prime, \Theta^\prime) }\cdot
\frac{ dn }{ dx_{+}dx_{-} }
\end{eqnarray}
we need to work out the Jacobian factor to move from
$(x_{+},x_{-})$ to $(\omega^\prime, \Theta^\prime)$
using the kinematical relations
\begin{eqnarray}
 x_{+}&=&\frac{1 - x_g (1-x_g/2)(1+\cos{\Theta^\prime})}
              {1 - x_g (1 + \cos{\Theta^\prime}) /2}
 \nonumber \\
 x_{-}&=&\frac{1-x_g}{1 - x_g (1 + \cos{\Theta^\prime}) /2}
 \label{jacobiankinematics}
\end{eqnarray}
with $x_g=2\omega^\prime/Q$.   
This can be done ecomically by 
noting $x_{-}=2-x_{+}-x_{g}$ and we obtain a simple formula
\begin{eqnarray}
J \equiv
\left|
\frac{ \partial(x_{+},x_{-}) }{ \partial(\omega^\prime,
\Theta^\prime) }\right|
=
\frac{2}{Q}
\left(
\frac{ \partial x_{-} }{ \partial \Theta^\prime }
\right)_{x_g}
=\frac{x_{-}^2 x_g \sin \Theta^\prime }{Q(1-x_g)}.   \label{jacobian}
\end{eqnarray}
The final form of $\wp_q(\omega^\prime,\Theta^\prime)$ from parent
quark is

\begin{equation}
\wp_q(\omega^\prime,\Theta^\prime)
= \frac{C_F}{4N_C}\gamma_0^2 
\frac{x_{-}^2 x_g \sin \Theta^\prime }{Q(1-x_g)}
\frac{x_{+}^2+x_{-}^2}{(1-x_{+})(1-x_{-})}.
\label{exactwpf}
\end{equation}
Using the kinematical relation
$
x_T^2=4(1-x_{+})(1-x_{-})(1-x_{g})/{x_{-}^2} \label{xt}
$
one finds the emission density
\begin{equation}
\wp_q(\omega^\prime,\Theta^\prime)
=
\frac{1}{\omega^\prime \sin{\Theta^\prime}}
\frac{C_F}{N_C}\gamma_0^2
~\frac{x_{+}^2  + x_{-}^2 }{2} 
\label{simplest}
%
\end{equation}

From the exact result (\ref{simplest}) we can easily
obtain the approximation for
small gluon transverse energy fraction $x_T$, i.e. for $x_-\to 1$,
\begin{eqnarray}
\wp_q(\omega^\prime,\Theta^\prime)
= \frac{ C_F }{ N_C }
\gamma_0^2 \frac{1}{\omega^\prime \sin \Theta^\prime}
~\frac{1 +(1-\frac{2\omega^\prime}{Q})^2}{ 2 }. \label{glapformula}
\end{eqnarray}
This formula contains the DGLAP splitting function
$
P_{qq}(z) = C_{F} \frac{1+z^2}{1-z}
$
as factor and can be seen to correspond to the expression
\begin{eqnarray}
\frac{ dn }{ dx_{+} dx_{T} } = \frac{ \alpha{_S} }{ \pi }
\frac{1}{ x_{T} } P_{qq}( x_{+} ),
\end{eqnarray}  
for, with
$x_{T} \equiv 2 \omega^\prime \sin \Theta^\prime /Q$, 
the appropriate Jacobian factor is
\begin{equation}
\left| \frac{\partial(x_{+},x_{T})}{\partial(\omega^\prime,
\Theta^\prime)} \right|
\approx \frac{2(1-x_{+})}{Q}.
\end{equation}  

The Eq. (\ref{glapformula}) differs from the DLA result by
the inclusion of the non-singular DGLAP term and the correct angular
dependence. We have computed  $w_q$ by
integrating the density as in (\ref{exactdoubleintegration}) 
{\it numerically} with the bounds
\begin{eqnarray}
   \frac{Q_0}{\sin \Theta} \le \omega^\prime \le P,
~~~\sin^{-1}\frac{Q_0}{\omega^\prime} \le \Theta^\prime \le \Theta.
\label{exactbound}
\end{eqnarray}
This yields an improved result for the gap probability calculated again as
in (\ref{fth}) or (\ref{fsym}) which we refer to as \lq\lq MLLA-n".

An {\it analytical} approximation (\lq\lq MLLA-a") can be obtained by simplifying
the integral over the 
nonsingular parts of the splitting function \cite{do}. One obtains
\begin{equation}
w_q=w_q^{DLA}-\frac{3}{4}\left[\ln\frac{Y+\lambda}{\lambda}
-e^\lambda\{E_1(\lambda)-E_1(Y+\lambda)\}\right].  
\label{wmlla}
\end{equation}
where $E_1(z)=\int_z^\infty dt e^{-t}/t$ and 
$w_q^{DLA}$ denotes the expression in Eq. (\ref{wdla}).
The results from these calculations will be discussed in the next
subsection. \\

\noindent
{\it Comparison of analytical results and a Monte Carlo calculation}\\
The result of our computations in  the different approximations are
displayed in Fig.~1. For the parameter $\lambda $ = 0.1 at
$\Lambda $ = 0.244~GeV we show the DLA result for the symmetric
gap (\ref{fsym}) and the MLLA results in the analytic approximation
(\ref{wmlla})
and in the numerical evaluation of (\ref{exactdoubleintegration}) 
with (\ref{glapformula}).
Also shown as data points are
the results from the ARIADNE Monte Carlo at the parton level
\cite{ariadnep} for the same parameters. This Monte-Carlo generates
the quark gluon cascade as a sequence of dipole emissions 
 and it is terminated as in the analytical
calculations by a transverse momentum cutoff \cite{ariadne}. 
Furthermore, there is no
limitation to the relative magnitude of the parameters (besides
$Q_0>\Lambda$). 

\begin{figure}
\begin{center}
\mbox{\epsfig{file=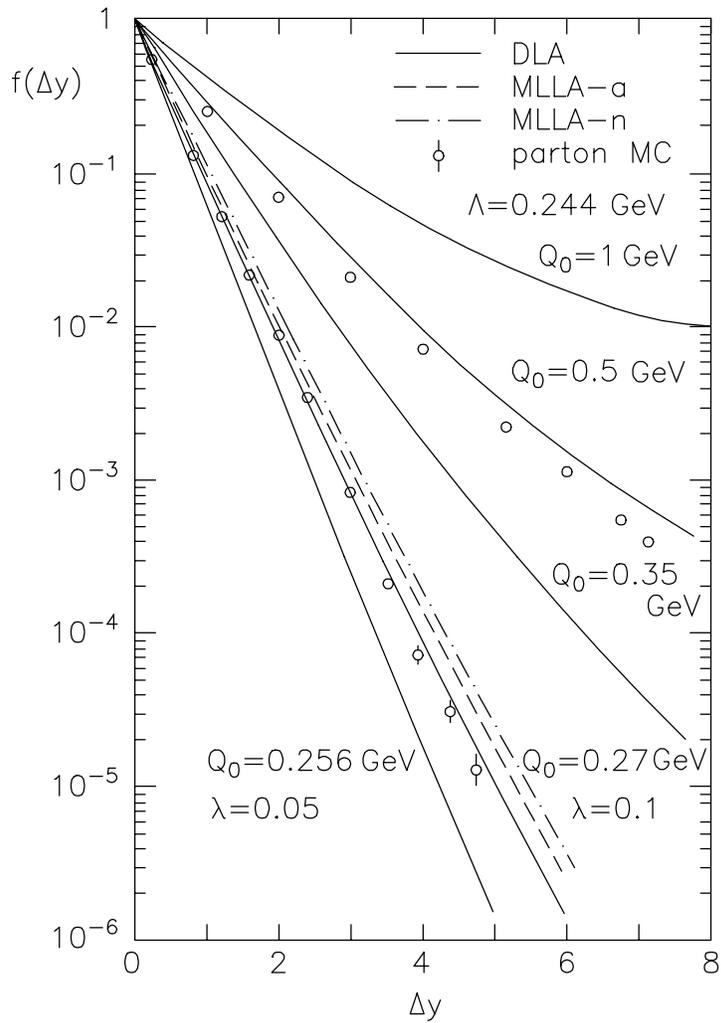,bbllx=2.5cm,bblly=5.6cm,%
bburx=16.7cm,bbury=27.cm,height=14.5cm}}
\end{center}

\caption
{Fraction of rapidity gap events in $e^+e^-$ annihilation
as a function of the full width of 
the symmetric gap. The DLA predictions from eq.~(\protect\ref{fsym})
for different values of the transverse momentum cutoff $Q_0$, and for
one value of $Q_0$ also the analytical (a) and the numerical (n) MLLA
results (see text). 
The data points refer to the calculations of the ARIADNE Monte Carlo
at the parton level for the corresponding parameters.}
\end{figure}

As can be seen from the figure the MLLA corrections are not very
large: about a factor or two after a decrease of the gap fraction
by 5 orders of magnitude. Of the same order is the deviation from
the parton Monte Carlo which takes into account effects not
considered here: the  correct angular recoil from the primary gluon
emission also for large angles 
and the spillover of secondary gluons into the gap if the
primary gluon is produced ouside the gap.
This effect could be responsible for the difference between the MC 
and the more accurate MLLA calculations. Surprisingly, the simple
DLA and the MC results are quite close to each other.

The figure also shows the very strong dependence on the cutoff
$Q_0$ for fixed $\Lambda $. Indeed, decreasing the
transverse momentum cutoff from $Q_0$ = 1~GeV down to
$Q_0$ = 0.27 the gap rate at $\Delta y$ = 6 decreases by 4
orders of magnitude. This comes from the singular emission
probability $dk_\perp/k_\perp \alpha_s(k_\perp)$ enhanced by the running
$\alpha_s$: the closer we come to the pole at $k_\perp=\Lambda $ the
more difficult it is to avoid the  gluon bremsstrahlung.

\section{Comparison with data}

Apparently the prediction for the gap probability in the
parton cascade depends sensitively on the parameter $Q_0$.
A determination of both parameters $\Lambda $ and $Q_0$ within
the MLLA-LPHD approach has been performed in the study of
particle energy spectra \cite{lo} applying the moment analysis
\cite{dkt}. This study has shown that the two parameters are
very close to each other and there is an upper limit to the
$\lambda$-parameter: 
\begin{equation}
\lambda \leq 0.1,  \label{lambda}
\end{equation}
whereas the absolute scale was found to be $Q_0 \simeq$
 270~MeV. A lower limit on $\lambda $ could in principle be obtained 
from the study of the mean multiplicity in a calculation of
yet higher accuracy. As central value for our prediction we
choose therefore $Q_0 $= 270~MeV and $\lambda $= 0.1
(or $\Lambda $ = 0.244~GeV) but smaller values of $\Lambda $
are allowed as well.

\begin{figure}
\begin{center}
\mbox{\epsfig{file=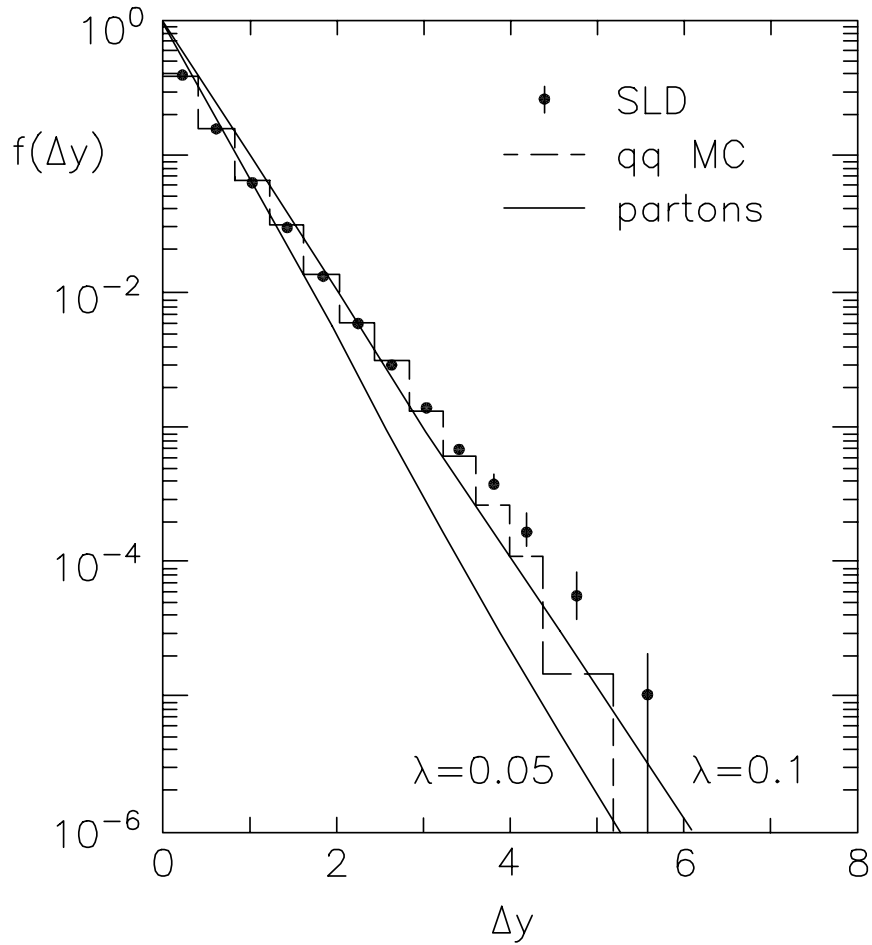,bbllx=3.0cm,bblly=9.0cm,%
bburx=16.5cm,bbury=23.7cm,height=13cm}}
\end{center}

\caption{Fraction of rapidity gap events in $e^+e^-$ annihilation
as a function of the full width of 
the symmetric gap. The data points refer to the measurement of 
gaps between charged particles ($\tau$-lepton events included) by
the SLD Collaboration \protect\cite{sld}. The dashed histogram shows the
expectation from the JETSET \protect\cite{jetset} Monte Carlo without
$\tau$-leptons. Also shown are the DLA predictions for the gaps of a parton
cascade as in Fig. 1.}
\end{figure}

In Fig.~2 we compare such results from the DLA (which agree closely
with those from the parton MC) with the experimental measurement of
the fraction of rapidity gap events from the SLD Collaboration
\cite{sld}. These data also contain a contribution from the
$\tau$-lepton events which is important for the large gaps
$\Delta y$. The data are well described by the JETSET Monte Carlo
\cite{jetset} including the $\tau$-decays. The Monte Carlo
predictions without $\tau$-decays are also shown in the figure
and indicate a steeper falloff for large $\Delta y$. Furthermore it
should be noted that
the data refer to rapidity gaps between charged particles. From
the Monte Carlo one would expect a slightly steeper dependence if
the gap is calculated from all final state particles.\cite{sld}

One can conclude from the figure that the parton cascade with the
low values for the scale parameters $Q_0$ and $\Lambda$ obtained
from previous analyses gives already enough suppression for
large rapidity gaps without any additional hadronization; values
$\lambda$ = 0.05-0.1 provide an adequate description of the gap
distribution (our predictions refer to gaps in the final state of all
charged and neutral particles).


\section{Discussion and further predictions}

The data are consistent with a description of the hadronic
final state in terms of a parton cascade 
terminated at a rather low cutoff of the order of the final state hadron
masses without any additional hadronization phase.
This could be interpreted in favour of
a dual correspondence of parton and hadron final states and
 a soft colour neutralization mechanism .

A good description of the data has been obtained \cite{sld} by the 
JETSET Monte Carlo 
\cite{jetset}. In this model the parton cascade is terminated at a cutoff
$Q_0 \sim $ O(1 GeV) and followed by a hadronization phase. According to
Fig. 1 the suppression of 
large rapidity gaps at a scale $Q_0 \sim 1$ GeV
is not so strong and most of the suppression must come from the resonance and
particle production in the hadronization phase.
In our parton calculation the strong suppression of large gaps again comes
mainly from the last stage of the cascade evolution where the radiation
becomes stronger because of the infrared enhancement  and the 
increasing coupling constant $\alpha_s$. It appears that the average
properties of the hadronization phase  can be well represented by the parton
bremsstrahlung cascade with running coupling.

In the following we note a few further predictions from the perturbative 
approach to the gap rate.
\begin{enumerate}
\item 
There is a difference in the gap fraction of quark and gluon jets
according to the respective colour factors in (\ref{wdla}).
This can be tested, for example,  by studying gaps in jets at high $p_T$ in
$p \bar p$ collisions between angles of, say, $\Theta_1 < \Theta_2 < 45^o$
using eq. (\ref{fth}).
The slope roughly doubles when going from quark to gluon jets.
This measurement would test the connection of the gap probability to single
gluon emission.

\item 
In the approximation considered the evolution of the parton cascades in
both hemispheres is independent, so the corresponding rates factorize.
For example the slope should double when going from the gap 
($\Theta_G,\frac{\pi}{2}$) to ($\Theta_G,\pi-\Theta_G$). 

\item 
The eq. (\ref{fth}) also predicts the energy dependence of the
slope. With rising jet energy the gap distribution gets slightly steeper:
at $Q$=200 GeV the slope is larger by 4 \% in comparison to $Q$=90 GeV.

\end{enumerate}
\noindent
It will be  interesting to study the gap events further in order
to test the proposed perturbative approach with soft colour
neutralization.

\section*{Acknowledgements}
We thank Sergio Lupia for useful discussions and help with  various problems 
of this work. One of us (W.O.) thanks the organizers of the workshop
L. Cifarelli, A. Kaidalov and V. A. Khoze for the invitation to this lively
workshop with good discussions to remember.

\end{document}